\documentclass[twocolumn,showpacs,aps,epsfig,nofootinbib]{revtex4}
\usepackage{graphicx}
\usepackage{amsfonts}
\usepackage{mathtools}
\usepackage{xcolor}

\usepackage{latexsym}
\usepackage{amssymb}
\usepackage{amssymb}
\usepackage[center]{subfigure}

\begin{document}	

	\newcommand{\bq}{\begin{equation}}
	\newcommand{\eq}{\end{equation}}
	\newcommand{\bqn}{\begin{eqnarray}}
	\newcommand{\eqn}{\end{eqnarray}}
	\newcommand{\nb}{\nonumber}
	\newcommand{\lb}{\label}
	\newcommand{\rr}{\bf r}
	\newcommand{\PRL}{Phys. Rev. Lett.}
	\newcommand{\PL}{Phys. Lett.}
	\newcommand{\PR}{Phys. Rev.}
	\newcommand{\CQG}{Class. Quantum Grav.}
	\newcommand{\hong}[1]{#1}
	
	\title{Existence of New Singularities in Einstein-Aether Theory}

	\author{R. Chan$^{1}$, M. F. A. da Silva$^{2}$ and V. H. Satheeshkumar$^{3}$}
		
	\email{chan@on.br, mfasnic@gmail.com, vhsatheeshkumar@gmail.com}
	
	\affiliation{
		$^{1}$Coordena\c{c}\~{a}o de Astronomia e Astrof\'{i}sica,
		Observat\'{o}rio Nacional (ON), Rio de Janeiro, RJ 20921-400, Brazil
		\\
		$^{2}$Departamento de F\'{i}sica Te\'{o}rica, 
		Universidade do Estado do Rio de Janeiro (UERJ), Rio de Janeiro, RJ 20550-900, Brazil
		\\
		$^{3}$Departamento de F\'{\i}sica, Universidade Federal do Estado do Rio de Janeiro (UNIRIO), Rio de Janeiro, RJ 22290-240, Brazil
	}
	
	\date{\today}
	
	\begin{abstract}
	How do the global properties of a Lorentzian manifold change when endowed with a vector field? 	This interesting question is  tackled in this paper within the framework of Einstein-Aether (EA) theory which has the most general diffeomorphism-invariant action involving a spacetime metric and a vector field. 	
	After classifying all the possible nine vacuum solutions with and without cosmological constant in Friedmann-Lema{\^{\i}}tre-Robertson-Walker (FLRW) cosmology, we show that there exist three singular solutions in the EA theory  which are not singular in the General Relativity (GR), all of them for $k=-1$, and another singular solution for $k=1$ in EA theory which does not exist in GR. This result is cross-verified by showing the focusing of timelike geodesics using the Raychaudhuri equation. These new singular solutions show that GR and EA theories can be completely different, even for the FLRW solutions when we go beyond flat geometry ($k=0$). In fact, they have different global structures. In the case where $\Lambda=0$ ($k=\pm 1$) the vector field defining the preferred direction is the unique source of the curvature.
	\end{abstract}
	
	\maketitle

\section{Introduction}

The theory of General Relativity (GR) underpins our best understanding of gravity. It has passed all the experimental tests, the first being the measurement of bending of light during the total solar eclipse of 1919 \cite{Crispino:2019yew} and the most recent confirmation come from  the direct detection of gravitational waves in 2015 \cite{Abbott:2016blz}, and the capture of black hole shadow image in 2019 \cite{Akiyama:2019cqa}. Despite the experimental success, a fundamental theoretical problem of GR is the existence of curvature singularities such as the ones in the big bang and black holes. A curvature singularity is a point of intense gravity where spacetime, and the laws of physics break down. Stephan Hawking and Roger Penrose proved that the curvature singularities are not artifacts of coordinates or symmetries but are inevitable under reasonable energy and causality conditions \cite{Hawking:1969sw}. Since then it has been clear that GR is not a final theory of gravitation and there have been several attempts to replace GR both at low and high energies \cite{Will:2018bme}. The Einstein-Aether (EA) theory belongs to a class of infrared theories.

In this paper, we are interested in studying how the timelike vector field dubbed aether in EA theory affects the nature of singularities in comparison to GR. In order to have a clear view, we first classify all the possible cosmological vacuum solutions in EA theory. Then, we compute the Kretschmann scalar to see if curvature singularities exist in both GR and EA theories. We found four solutions that are non-singular (and one of them is non-existing) in GR theory but are singular in EA theory, three of them within the experimentally allowed parameter space. We reconfirm this by studying the focusing of congruence of timelike geodesics using the Raychaudhuri equation. It is noteworthy to mention that recently, in EA theory, we stumbled upon an interesting case of a new singularity in FLRW solution for a dark energy fluid with equation of state $p=-\rho$, with constant energy density \cite{Campista:2018gfi}. 
Even in GR, with a rather exotic and unphysical matter, one can get physical singularities that are unaccompanied by geodesic incompleteness which are called ``sudden'' singularities \cite{Barrow:2004xh} \cite{Barrow:2004hk} \cite{FernandezJambrina:2004yy}. A large class of such singularities were also found in a simple Friedmann cosmology containing only a scalar-field with a power-law self-interaction potential \cite{Barrow:2015sga}. 
The paper is organized as follows.  The Section II presents the quick overview of EA field equations.
In Section III, we classify the vacuum cosmological solutions and compare them with the corresponding cases in GR theory. We also prove the existence of curvature singularities by computing the Kretschmann curvature scalar invariant.  And finally we end with a summary and conclusion in \textcolor{black}{Section IV}.

\section{Einstein-Aether theory}

The EA theory in the current form was introduced in 2001 to study the preferred frame effects in gravitation and cosmology \cite{Jacobson:2000xp}. In this  generally  covariant  theory,  the  local  Lorentz  Invariance is  broken  by  a  dynamical  unit timelike vector field $u^a$ often referred to as aether.  The general action of the EA theory is given by, 
\bq S = \int \sqrt{-g}~(L_{\rm Einstein}+L_{\rm aether}+L_{\rm matter}) d^{4}x,
\label{action} 
\eq
where, 
\bqn 
L_{\rm Einstein} &=&  \frac{1}{16\pi G} \left( R - 2 \Lambda \right), \\ 
L_{\rm aether} &=&  \frac{1}{16\pi G} [-K^{ab}{}_{mn} \nabla_a u^m \nabla_b u^n \nb \\
&& ~ ~ ~~~~~~ + \lambda(g_{ab}u^a u^b + 1)],
\lb{LEAG}
\eqn
with
\bq {{K^{ab}}_{mn}} = c_1 g^{ab}g_{mn}+c_2\delta^{a}_{m} \delta^{b}_{n}
+c_3\delta^{a}_{n}\delta^{b}_{m}-c_4u^a u^b g_{mn},
\lb{Kab}
\eq
and the $c_i$ being dimensionless coupling constants, and $\lambda$
a Lagrange multiplier enforcing the unit timelike constraint on the aether. 

The last term, $L_{\rm matter}$ is the matter Lagrangian.

In the weak-field, slow-motion limit EA theory reduces to Newtonian gravity with a value of  Newton's constant $G_{\rm N}$ related to the parameter $G$ in the action (\ref{action})  by  {\cite{Garfinkle}},
\bq
G = G_N\left(1-\frac{c_1+c_4}{2}\right).
\lb{Ge}
\eq
The coupling constant $G$ of EA theory is equal to the usual Newtonian gravitational constant $G_N$ for $c_1 = -c_4$ and not necessarily $c_1 = c_4 = 0$. The Newtonian limit is recovered only for $c_1 + c_4 < 2$. If $c_1 + c_4 > 2$ gravity is repulsive, while for $c_1 + c_4 = 2$ the coupling  constant $G$ is zero, which means that there is no coupling between gravity and matter in this theory \cite{Jacobson:2008aj}.

The field equations, obtained by extremizing the action with respect to the independent  variables $\lambda$, $u^a$ and $g_{mn}$ are given by \cite{Garfinkle},
\bq
g_{ab}u^a u^b = -1,
\label{LagMul}
\eq
\bq
\nabla_a \left( K^{am}_{bn} \nabla_m u^n, \right) + c_4 u^m \nabla_m u_n \nabla_b u^n + \lambda u_b = 0,
\eq
\bq
G^{Einstein}_{ab} = T^{aether}_{ab} +8 \pi G \, T^{matter}_{ab},
\label{EA}
\eq
with 
\bqn
G^{Einstein}_{ab} &=& R_{ab} - \frac{1}{2} R\, g_{ab}  + \Lambda \, g_{ab}, \\
T^{\, aether}_{ab}&=& \nabla_c [ J^c\;_{(a} u_{b)} + u^c J_{(ab)} - J_{(a} \;^c u_{b)}]  \nb\\
&& - \frac{1}{2} g_{ab} J^c_d \nabla_c u^d+ \lambda u_a u_b  \nb \\
& & + c_1 [\nabla_a u_c \nabla_b u^c - \nabla^c u_a \nabla_c u_b] \nb \\
&& + c_4 \, u^m \nabla_m u_a u^n \nabla_n u_a,\\
T^{\, matter}_{ab} &=&  \frac{- 2}{\sqrt{-g}} \frac{\delta \left( \sqrt{-g} L_{matter} \right)}{\delta g_{ab}}.
\label{fieldeqs}
\eqn

By demanding the ghost-free condition of the tensor, vector and scalar parts of the linearly perturbed AE action, we get the following respective constraints on the free parameters $c_i$ \cite{Foster:2005dk}.
$$c_1+ c_3 < 1$$
$$c_1+ c_4 > 0$$
$$c_2 > -1$$
Thus theoretical consistency requires $\beta = c_1+3c_2+c_3 > -2$, which we will use later. These free parameters  have been severely constrained using many observational/experimental tests such as the primordial nucleosynthesis \cite{Carroll:2004ai}, ultra-high energy cosmic rays \cite{Elliott:2005va}, the solar system tests \cite{Eling:2003rd, GrEAsser:2005bg}, binary pulsars \cite{Foster:2007gr, Yagi:2013ava}, and more recently gravitational waves \cite{Gong:2018cgj, Oost:2018tcv}.

\section{Classification of Cosmological Models}

The most general isotropic and homogeneous universe is described by a FLRW metric,
\bq
ds^2= -dt^2+B(t)^2\left[\frac{dr^2}{1-kr^2} +r^2 d\theta^2 +r^2 \sin^2 \theta d\phi^2\right],
\lb{ds2}
\eq
where, $B(t)$ is the scale factor {and} $k$ is a Gaussian curvature the space at a given time. According to observations by WMAP and Planck experiments, this metric is a good description of our universe as it is spatially homogeneous and isotropic when averaged over large scales. This leaves us with a choice of
\bq
u^a=(1,0,0,0).
\eq

The standard definitions of the Hubble parameter $H(t)$, the deceleration parameter $q(t)$ and the redshift are given by, respectively,
\bqn
H(t) &=& \frac{\dot{B(t)}}{B(t)},\\ 
\nb \\
q(t) &=& - \frac{\ddot{B(t)} B(t)}{\dot{B(t)}^2},
\eqn
where the symbol dot denotes the differentiation with respect to the time coordinate.

The Friedmann-Lema\^{i}tre equations are given by,
\bqn
\left( 1 + \frac{\beta}{2}  \right) \left(  \frac{\dot{B(t)}}{B(t)} \right)^2   &=& \frac{\Lambda}{3}  - \frac{k}{B(t)^2}, \\ 
\nb \\
\left( 1 + \frac{\beta}{2}  \right)   \frac{\ddot{B(t)}}{B(t)}   &=& \frac{\Lambda}{3}, 
\eqn
where $\Lambda$ is the cosmological constant. For recent literature on EA cosmology, the reader may consult the references \cite{Coley:2019tyx}, \cite{Leon:2019jnu} and \cite{Trinh:2018pcb}.
	 
Since we are interested in the curvature singularities, an important quantity to be computed is the Kretschmann scalar. For FLRW metric, it is  given by
	\bq
	K   = \frac{12}{B^4} \left[k^2+2 k \dot B(t)^2+\dot B(t)^4+\ddot B(t)^2 B(t)^2\right].
	\lb{K}
	\eq

%
A singularity always implies focusing of geodesics, although focusing alone cannot imply a singularity as pointed out by Landau \cite{Landau:1982dva}. Having already established the presence of singularities, we now use the focusing of timelike geodesics to reinforce our results. For a similar analysis, see \cite{Albareti:2012se}. The expansion rate of the congruence of geodesics as seen by comoving observers is given by the Raychaudhuri equation \cite{Raychaudhuri:1953yv}\cite{Hawking:1973uf}\cite{Wald:1984}\cite{Senovilla:2006},
\bq
\frac{d\theta}{d\tau}= - \frac{\theta^2}{3}-\sigma_{\mu\nu}\sigma^{\mu\nu}+w_{\mu\nu}w^{\mu\nu} -R_{\mu\nu}k^{\mu}k^{\nu},
\label{Raychaudhuri}
\eq
where $\theta,\sigma^{\mu\nu},w^{\mu\nu}$ are respectively the expansion, shear and twist of the congruence of geodesics, and $\tau$ is the proper time along a geodesic with a tangent vector field,  $k^{\mu}=dx^{\mu}/d\tau$. The Raychaudhuri equation has geometrical meaning and has no connection \textit{a priori} to the gravitational theory which only enters through the term  $-R_{\mu\nu}\xi^{\mu}\xi^{\nu}$.
For the FLRW metric, ($\tau=t$), assuming the vector $k^{\mu}=\delta^\mu_t$ as the four-velocity, both shear $\sigma^{\mu\nu}$ and twist $w^{\mu\nu}$ are zero, while the expansion and curvature terms for timelike geodesics are given by,
\bqn
\theta &=& \frac{3}{2}\frac{{\dot B}(t)}{B(t)},
\\
-R_{\mu\nu}k^{\mu}k^{\nu} &=& -R_{tt}, 
\eqn
where $R_{tt}$ is the component $tt$ of the Ricci tensor.
Thus, the expansion rate of congruence of timelike geodesics is given by
\bq
\frac{d\theta}{dt}= -\frac{3}{4}\left[\frac{\dot B(t)}{B(t)}\right]^2 +3\frac{\ddot B(t)}{B(t)}=-3 H^2\left(q+\frac{1}{4}\right).
\eq
In the next sections we will calculate the Kretschmann scalar and the expansion rate for all the obtained solutions (for both $B_1(t)$ and $B_2(t)$, except when they are static or imaginary).  Only for calculating the focussing of geodesics, we have used the the solutions before normalizations 
for the sake of simplicity of the analysis.

\subsection{{Solutions for $\Lambda > 0$}}

\subsubsection{{$\textbf{k = 1}$}}

\bigskip
The Friedmann-Lema\^{i}tre equations yield the following two solutions for $\Lambda > 0, k = 1$,
{
\bqn
B_1(t)&&=\frac{1}{2\sqrt{\Lambda (\beta+2)}} \times \nb \\
&&\left[e^{\sqrt{\frac{2\Lambda}{3(\beta+2)}}(t_0-t)}+3\left(\beta+2\right)e^{-\sqrt{\frac{2\Lambda}{3(\beta+2})} (t_0-t)}\right],
\eqn
\bqn
B_2(t)&&=\frac{1}{2\sqrt{\Lambda (\beta+2)}}\times \nb \\
&&\left[e^{\sqrt{\frac{2\Lambda}{3(\beta+2})}(t-t_0)}+3\left(\beta+2\right)e^{- \sqrt{\frac{2\Lambda}{3(\beta+2})} (t-t_0)}\right].
\eqn
Note that there is an integration constant $t_0$ which must be chosen in a such way that we have the Sitter solutions \cite{d'Inverno1992} at the limit of GR theory. Thus,
\bq
t_0=\frac {\ln{ 6 }}{2}\sqrt{\frac{3}{\Lambda}}.
\lb{t0}
\eq
As usual the $t_0$ may be the current time where the scale factor is normalized $B(t_0)=1$.
Thus,

\bqn
B1(t) &=& \frac{1}{2 \sqrt{\Lambda}} \left[
  \frac{3\, e^{\sqrt{\frac{2\Lambda}{3(\beta+2)}} (t_0-t)}}{\sqrt{\Lambda} + \epsilon \sqrt{\Lambda - 3}} +
  \frac{\sqrt{\Lambda} + \epsilon \sqrt{\Lambda - 3}}{e^{\sqrt{\frac{2\Lambda}{3(\beta+2)}} (t_0-t)}} 
   \right],\\ 
\nb \\
B_2(t) &=& \frac{1}{2 \sqrt{\Lambda}} \left[  
\frac{3\, e^{\sqrt{\frac{2\Lambda}{3(\beta+2)}} (t-t_0)}}{\sqrt{\Lambda} + \epsilon \sqrt{\Lambda - 3}} +
\frac{\sqrt{\Lambda} + \epsilon \sqrt{\Lambda - 3}}{e^{\sqrt{\frac{2\Lambda}{3(\beta+2)}} (t-t_0)}} 
\right],
\eqn
where $\epsilon=\pm 1$.}
The solution only exists for $\Lambda - 3 > 0$ and $\beta+2 > 0$, {including $\beta=0$}.

The Hubble parameter $H(t)$ and the deceleration parameter $q(t)$ corresponding to this metric are given by
\bqn
H_1(t) &=&  \sqrt{\frac{2\Lambda}{3(\beta+2)}} \nb \\
&& \times \left[  \frac{-3\, e^{\sqrt{\frac{2\Lambda}{3(\beta+2)}} 2(t_0-t)} + 
	\left(  \sqrt{\Lambda} + \epsilon \sqrt{\Lambda - 3} \right)^2 }
{ 3\, e^{\sqrt{\frac{2\Lambda}{3(\beta+2)}} 2(t_0-t)} + \left(  \sqrt{\Lambda} + \epsilon \sqrt{\Lambda - 3} \right)^2 } 
\right], \\
H_2(t) &=& - \sqrt{\frac{2\Lambda}{3(\beta+2)}} \nb \\
&&  \times \left[  \frac{ -3\, e^{\sqrt{\frac{2\Lambda}{3(\beta+2)}} 2(t-t_0)} + 
	\left(  \sqrt{\Lambda} + \epsilon \sqrt{\Lambda - 3} \right)^2 }
{ 3\, e^{\sqrt{\frac{2\Lambda}{3(\beta+2)}} 2(t-t_0)} + \left(  \sqrt{\Lambda} + \epsilon \sqrt{\Lambda - 3} \right)^2 } 
\right], \\
\nb \\
q_1(t) &=& - \left[  \frac{ 3\, e^{\sqrt{\frac{2\Lambda}{3(\beta+2)}} 2(t_0-t)} + 
	\left(  \sqrt{\Lambda} + \epsilon \sqrt{\Lambda - 3} \right)^2 }
{ -3\, e^{\sqrt{\frac{2\Lambda}{3(\beta+2)}} 2(t_0-t)} + 
	\left(  \sqrt{\Lambda} + \epsilon \sqrt{\Lambda - 3} \right)^2 } 
\right]^2, \\
\nb \\
q_2(t) &=& - \left[  \frac{ 3\, e^{\sqrt{\frac{2\Lambda}{3(\beta+2)}} 2(t-t_0)} + 
	\left(  \sqrt{\Lambda} + \epsilon \sqrt{\Lambda - 3} \right)^2 }
{ -3\, e^{\sqrt{\frac{2\Lambda}{3(\beta+2)}} 2(t-t_0)} + 
	\left(  \sqrt{\Lambda} + \epsilon \sqrt{\Lambda - 3} \right)^2 } 
\right]^2. 
\eqn
This metric is non-singular for all values of $\beta + 2 > 0$.

The Raychaudhuri equation is given by
\bqn
\frac{d\theta}{d\tau}(B_1)&&= \frac{3\Lambda}{2(\beta+2)}\frac{1}{\left[ 3(\beta+2)+e^{\frac{2}{3}\sqrt{\frac{6\Lambda}{\beta+2}}(-t+C1)}\right]^2}\times\nb\\
&& \left[e^{\frac{4}{3}\sqrt{\frac{6\Lambda}{\beta+2}}(-t+C1)}+10(\beta+2) e^{\frac{2}{3}\sqrt{\frac{6\Lambda}{\beta+2}}(-t+C1)}+\right.\nb\\
&&\left. 9(\beta+2)^2\right],
\lb{Lambda>0k1B1}
\eqn

\bqn
\frac{d\theta}{d\tau}(B_2)&&= \frac{3\Lambda}{2(\beta+2)}\frac{1}{\left[ 3(\beta+2)+e^{-\frac{2}{3}\sqrt{\frac{6\Lambda}{\beta+2}}(-t+C2)}\right]^2}\times\nb\\
&& \left[e^{-\frac{4}{3}\sqrt{\frac{6\Lambda}{\beta+2}}(-t+C2)}+10(\beta+2) e^{-\frac{2}{3}\sqrt{\frac{6\Lambda}{\beta+2}}(-t+C2)}+\right.\nb\\
&&\left. 9(\beta+2)^2\right],
\lb{Lambda>0k1B2}
\eqn
where $C1$ and $C2$ are, hereinafter, integration constants.

\subsubsection{{$\textbf{k = 0}$}}

\bigskip
The Friedmann-Lema\^{i}tre equations yield the following two solutions for $\Lambda > 0, k = 0$,

\bq
B_1(t)=e^{\sqrt{\frac{2\Lambda}{3(\beta+2)}} (t_0-t)},
\eq
\bq
B_2(t)=e^{\sqrt{\frac{2\Lambda}{3(\beta+2)}} (t-t_0)}.
\eq
Note again that the integration constant $t_0$, given by equation (\ref{t0}), must be chosen in a such way that we have the Sitter solutions \cite{d'Inverno1992} at the limit of GR theory. Thus,
In the usual normalization $B(t_0)=1$ gives the same solutions.

The solution only exists for $\beta+2 > 0$.

The Hubble parameter $H(t)$ and the deceleration parameter $q(t)$ corresponding to the cases of this metric are given by
\bqn
H_1(t) &=& -\sqrt{\frac{2 \Lambda}{3(\beta+2)}},\\ 
\nb \\
H_2(t) &=& \sqrt{\frac{2 \Lambda}{3(\beta+2)}}, \\
\nb \\
q_1(t) &=& - 1,\\ 
\nb \\
q_2(t) &=& - 1.
\eqn
This metric is non-singular for all values of $\beta + 2 > 0$ {since the Kretschmann
scalar is a constant ($K=\frac{32}{3} \frac{ \Lambda^2}{(\beta+2)^2}$)}.

The Raychaudhuri equation is given by
\bqn
\frac{d\theta}{d\tau}(B_1)&=& \frac{3}{2}\frac{\Lambda}{\beta+2},
\lb{Lambda>0k0B1}
\eqn
\bqn
\frac{d\theta}{d\tau}(B_2)&=& \frac{3}{2}\frac{\Lambda}{\beta+2}.
\lb{Lambda>0k0B2}
\eqn

\subsubsection{{$\textbf{k = -1}$}}

\bigskip
The Friedmann-Lema\^{i}tre equations yield the following two solutions for $\Lambda > 0, k = -1$.
\bqn
B_1(t)&=&\frac{1}{2\sqrt{\Lambda (\beta+2)}} \times \nb \\
&&\left[e^{\sqrt{\frac{2\Lambda}{3(\beta+2)}}(t_0-t)}-3\left(\beta+2\right)e^{-\sqrt{\frac{2\Lambda}{3(\beta+2)}} (t_0-t)}\right],
\eqn
\bqn
B_2(t)&=&\frac{1}{2\sqrt{\Lambda (\beta+2)}}\times \nb \\
&&\left[e^{\sqrt{\frac{2\Lambda}{3(\beta+2)}}(t-t_0)}-3\left(\beta+2\right)e^{- \sqrt{\frac{2\Lambda}{3(\beta+2})} (t-t_0)}\right].
\eqn
As before the integration constant $t_0$ given by equation (\ref{t0}), it is chosen in a such way that we have the Sitter solutions \cite{d'Inverno1992} at the limit of GR theory. Thus,
as usual the $t_0$ may be the current time where the scale factor is normalized $B(t_0)=1$.
Thus,
\bqn
B_1(t) &=& \frac{1}{2 \sqrt{\Lambda}} \left[ -\frac{3\, e^{\sqrt{\frac{2\Lambda}{3(\beta+2)}} (t_0-t)}}{ \epsilon \sqrt{\Lambda + 3} + \sqrt{\Lambda}}  + \frac{ \epsilon \sqrt{\Lambda + 3} + \sqrt{\Lambda}}{e^{\sqrt{\frac{2\Lambda}{3(\beta+2)}} (t_0-t)}} \right],\\ 
\nb \\
B_2(t) &=& \frac{1}{2 \sqrt{\Lambda}} \left[ -\frac{3\, e^{\sqrt{\frac{2\Lambda}{3(\beta+2)}} (t-t_0)}}{ \epsilon \sqrt{\Lambda + 3} + \sqrt{\Lambda}}  + \frac{\epsilon \sqrt{\Lambda + 3} + \sqrt{\Lambda}}{e^{\sqrt{\frac{2\Lambda}{3(\beta+2)}} (t-t_0)}} \right].
\eqn
The solution only exists for $\beta+2 > 0$.

The Hubble parameter $H(t)$ and the deceleration parameter $q(t)$ corresponding to this metric are given by
\bqn
H_1(t) &=& -\sqrt{\frac{2\Lambda}{3(\beta+2)}} \nb \\
&& \times \left[  \frac{ 3\, e^{\sqrt{\frac{2\Lambda}{3(\beta+2)}} 2(t_0-t)} + \left( \epsilon \sqrt{\Lambda + 3 }+ \sqrt{\Lambda} \right)^2 }{ 3\, e^{\sqrt{\frac{2\Lambda}{3(\beta+2)}} 2(t_0-t)} - \left(  \epsilon \sqrt{\Lambda + 3 }+ \sqrt{\Lambda} \right)^2 } \right], \\
\nb \\
H_2(t) &=& \sqrt{\frac{2\Lambda}{3(\beta+2)}} \nb \\
&& \times  \left[  \frac{ 3\, e^{\sqrt{\frac{2\Lambda}{3(\beta+2)}} 2(t-t_0)} + \left(  \epsilon \sqrt{\Lambda + 3 } + \sqrt{\Lambda} \right)^2 }{ 3\, e^{\sqrt{\frac{2\Lambda}{3(\beta+2)}} 2(t-t_0)} - \left(  \epsilon \sqrt{\Lambda + 3 } + \sqrt{\Lambda} \right)^2 } \right],  \\
\nb \\
q_1(t) &=& - \left[  \frac{ 3\, e^{\sqrt{\frac{2\Lambda}{3(\beta+2)}} 2(t_0-t)} - \left(  \epsilon \sqrt{\Lambda + 3 } + \sqrt{\Lambda} \right)^2 }{ 3\, e^{\sqrt{\frac{2\Lambda}{3(\beta+2)}} 2(t_0-t)} + \left(  \epsilon \sqrt{\Lambda + 3 }+ \sqrt{\Lambda} \right)^2 } \right]^2,\\ 
\nb \\
q_2(t) &=& - \left[  \frac{ 3\, e^{\sqrt{\frac{2\Lambda}{3(\beta+2)}} 2(t-t_0)} - \left( \epsilon \sqrt{\Lambda + 3 } + \sqrt{\Lambda} \right)^2 }{ 3\, e^{\sqrt{\frac{2\Lambda}{3(\beta+2)}} 2(t-t_0)} + \left( \epsilon \sqrt{\Lambda + 3 } + \sqrt{\Lambda} \right)^2 } \right]^2.
\eqn

The Kretschmann scalar for the metric is singular at
\bqn
t_{sing}(B_1) &&= t_0 - \nb \\
&&\sqrt{\frac{3(\beta+2)}{8\Lambda}} \ln \left[ \frac{2}{3}\Lambda + \frac{2}{3}\epsilon \sqrt{\Lambda (\Lambda+3)}+1 \right], \\ 
\nb \\
t_{sing}(B_2) &&= t_0 + \nb \\
&&\sqrt{\frac{3(\beta+2)}{8\Lambda}} \ln \left[\frac{2}{3}\Lambda + \frac{2}{3}\epsilon \sqrt{\Lambda (\Lambda+3)}+1 \right].
\eqn

The metric is not singular for $\beta = 0$ since in this time the curvature invariant being $\frac{8 \Lambda^2}{3}$. But, the metric is singular for $\beta + 2 > 0$ 
with $\beta \neq 0$. This means it is a new singularity.

The Raychaudhuri equation is given by
\bqn
\frac{d\theta}{d\tau}(B_1)&&= \frac{3\Lambda}{2(\beta+2)}\frac{1}{\left[ -3(\beta+2)+e^{\frac{2}{3}\sqrt{\frac{6\Lambda}{\beta+2}}(-t+C1)}\right]^2}\times\nb\\
&& \left[e^{\frac{4}{3}\sqrt{\frac{6\Lambda}{\beta+2}}(-t+C1)}-10(\beta+2) e^{\frac{2}{3}\sqrt{\frac{6\Lambda}{\beta+2}}(-t+C1)}+\right.\nb\\
&&\left.9(\beta+2)^2\right],
\lb{Lambda>0k-1B1}
\eqn
\bqn
\frac{d\theta}{d\tau}(B_2)&&= \frac{3\Lambda}{2(\beta+2)}\frac{1}{\left[ -(3\beta+2)+e^{-\frac{2}{3}\sqrt{\frac{6\Lambda}{\beta+2}}(-t+C2)}\right]^2}\times\nb\\
&& \left[e^{-\frac{4}{3}\sqrt{\frac{6\Lambda}{\beta+2}}(-t+C2)}-10(\beta+2) e^{-\frac{2}{3}\sqrt{\frac{6\Lambda}{\beta+2}}(-t+C2)}+\right.\nb\\
&&\left. 9(\beta+2)^2\right].
\lb{Lambda>0k-1B2}
\eqn

\subsection{{Solutions for $\Lambda = 0$}}

\subsubsection{{$\textbf{k = 1}$}}

\bigskip
The Friedmann-Lema\^{i}tre equations yield the following two solutions for $\Lambda = 0, k = 1$,

\bqn
B_1(t) &=& 1 + \sqrt{\frac{-2}{2 +  \beta}} (t_0 - t), \\ 
\nb \\
B_2(t) &=& 1 + \sqrt{\frac{-2}{2 +  \beta}} (t - t_0).
\eqn
The solution only exists for $\beta+2 < 0$.

The Hubble parameter $H(t)$ and the deceleration parameter $q(t)$ corresponding to this metric are given by
\bqn
H_1(t) &=& - \left[ \sqrt{\frac{2 +  \beta}{-2}} +  (t_0 - t) \right]^{-1}, \\ 
\nb \\
H_2(t) &=& \left[ \sqrt{\frac{2 +  \beta}{-2}} +  (t - t_0) \right]^{-1}, \\
\nb \\
q_1(t) &=& 0,\\ 
\nb \\
q_2(t) &=& 0.
\eqn

The Kretschmann scalar for the metric is singular at
\bqn
t_{sing}(B_1) &=& t_0 + \sqrt{\frac{2 +  \beta}{-2}}, \\ 
\nb \\
t_{sing}(B_2) &=& t_0 - \sqrt{\frac{2 +  \beta}{-2}} .
\eqn
For $\beta = 0$, the solution does not exist. That means it is a new singularity but exists for EA theory.

The Raychaudhuri equation is given by
\bqn
\frac{d\theta}{d\tau}(B_1)&=& -\frac{3}{\left[-2 t+C1\sqrt{-2(\beta+2)}\right]^2},
\lb{Lambda=0k1B1}
\eqn
\bqn
\frac{d\theta}{d\tau}(B_2)&=& -\frac{3}{\left[-2 t+C2\sqrt{-2(\beta+2)}\right]^2}.
\lb{Lambda=0k1B2}
\eqn

\subsubsection{{$\textbf{k = 0}$}}

\bigskip
The Friedmann-Lema\^{i}tre equations yield the following solution for $\Lambda = 0, k = 0$,
\bqn
B(t) &=& 1,
\eqn

The Hubble parameter $H(t)$ and the deceleration parameter $q(t)$ corresponding to this metric are given by
\bqn
H(t) &=& 0,
\nb \\
q(t) &=& 0.
\eqn
This metric is never singular and independent of the value of $\beta$.

\subsubsection{{$\textbf{k = -1}$}}

\bigskip
The Friedmann-Lema\^{i}tre equations yield the following two solutions for $\Lambda = 0, k = -1$,
\bqn
B_1(t) &=& 1 + \sqrt{\frac{2}{2 +  \beta}} (t_0 - t) \\ 
\nb \\
B_2(t) &=& 1 + \sqrt{\frac{2}{2 +  \beta}} (t - t_0)
\eqn
The solution only exists for $\beta+2 > 0$.

The Hubble parameter $H(t)$ and the deceleration parameter $q(t)$ corresponding to this metric are given by
\bqn
H_1(t) &=& - \left[ \sqrt{\frac{2 +  \beta}{2}} +  (t_0 - t) \right]^{-1}, \\ 
\nb \\
H_2(t) &=& \left[ \sqrt{\frac{2 +  \beta}{2}} +  (t - t_0) \right]^{-1}, \\
\nb \\
q_1(t) &=& 0,\\ 
\nb \\
q_2(t) &=& 0.
\eqn

The Kretschmann scalar for the metric is singular at
\bqn
t_{sing}(B_1) &=& t_0 + \sqrt{\frac{2 +  \beta}{2}}, \\ 
\nb \\
t_{sing}(B_2) &=& t_0 - \sqrt{\frac{2 +  \beta}{2}}. 
\eqn
For $\beta = 0$ which corresponds to GR, the solution exists and is never singular with the curvature invariant being null. But, the metric is singular for $\beta + 2 > 0$ such that $\beta \neq 0$. This means it is a new singularity.

The Raychaudhuri equation is given by
\bqn
\frac{d\theta}{d\tau}(B_1)&=& -\frac{3}{\left[-2 t+C1\sqrt{2(\beta+2)}\right]^2},
\lb{Lambda=0k-1B1}
\eqn
\bqn
\frac{d\theta}{d\tau}(B_2)&=& -\frac{3}{\left[-2 t+C2\sqrt{2(\beta+2)}\right]^2}.
\lb{Lambda=0k-1B2}
\eqn

\subsection{{Solutions for $\Lambda < 0$}}

\subsubsection{{$\textbf{k = 1}$}}

\bigskip
The Friedmann-Lema\^{i}tre equations yield the following two solutions for $\Lambda < 0, k = 1$,
\bqn
B_1(t) = \sqrt{-\frac{3}{|\Lambda|}}
\sin{\left[ \sqrt{\frac{2 |\Lambda|}{3(\beta+2)}} (t_0-t) \right]}, \\ 
\nb \\
B_2(t) =  \sqrt{-\frac{3}{|\Lambda|}}
\sin{\left[ \sqrt{\frac{2 |\Lambda|}{3(\beta+2)}}  (t-t_0) \right]}.
\eqn
These solutions are imaginaries and substituting into the Friedmann equations we can show the solutions only exist
for $\Lambda=3$, thus there are no solutions for this case, since $\Lambda<0$.

\subsubsection{{$\textbf{k = 0}$}}

\bigskip
The Friedmann-Lema\^{i}tre equations yield the following two solutions for $\Lambda < 0, k = 0$,
\bqn
B_1(t) &=&  e^{\sqrt{\frac{-2 |\Lambda|}{3(\beta+2)}} (t-t_0)} \\ 
\nb \\
B_2(t) &=&  e^{\sqrt{\frac{-2 |\Lambda|}{3(\beta+2)}} (t_0-t)}
\eqn
{The solution only exists for $\beta+2 < 0$.}

The Hubble parameter $H(t)$ and the deceleration parameter $q(t)$ corresponding to this metric are given by
\bqn
H_1(t) &=& \sqrt{\frac{-2 |\Lambda|}{3(\beta+2)}}, \\ 
\nb \\
H_2(t) &=& - \sqrt{\frac{-2 |\Lambda|}{3(\beta+2)}},  \\
\nb \\
q_1(t) &=&  - 1,\\ 
\nb \\
q_2(t) &=&  - 1.
\eqn
For $\beta = 0$, the solution does not exist. The solution exists and is non-singular only for EA theory.

The Raychaudhuri equation is given by
\bqn
\frac{d\theta}{d\tau}(B_1)&=& -\frac{3}{2} \frac{\lvert \Lambda \rvert }{\beta+2},
\lb{Lambda<0k0B1}
\eqn
\bqn
\frac{d\theta}{d\tau}(B_2)&=&  -\frac{3}{2} \frac{\lvert \Lambda \rvert }{\beta+2}.
\lb{Lambda<0k0B2}
\eqn

\subsubsection{{$\textbf{k = -1}$}}

\bigskip
The Friedmann-Lema\^{i}tre equations yield the following two solutions for $\Lambda < 0, k = -1$,
\bqn
B_1(t) =  \sqrt{\frac{3}{|\Lambda|}} \sin{\left[ \sqrt{\frac{2 |\Lambda|}{3(\beta+2)}}  (t-t_0) \right]} 
\nb \\
B_2(t) = \sqrt{\frac{3}{|\Lambda|}} \sin{\left[ \sqrt{\frac{2 |\Lambda|}{3(\beta+2)}}  (t_0-t) \right]}\,\,\,\,\,\,\,
\eqn
In the usual normalization $B(t_0)=1$ we have
\bqn
B_1(t) &=&  \sqrt{\frac{3}{|\Lambda|}} \sin{\left( \sqrt{\frac{2 |\Lambda|}{3(\beta+2)}}  (t-t_0) + \sin^{-1}{\sqrt{\frac{|\Lambda|}{3}}} \right)} \,\,\,\,\,\,\,\\ 
\nb \\
B_2(t) &=&  \sqrt{\frac{3}{|\Lambda|}} \sin{\left( \sqrt{\frac{2 |\Lambda|}{3(\beta+2)}}  (t_0-t) + \sin^{-1}{\sqrt{\frac{|\Lambda|}{3}}} \right)}\,\,\,\,\,\,\,
\eqn
The solution only exists for $\beta+2 > 0$.

The Hubble parameter $H(t)$ and the deceleration parameter $q(t)$ corresponding to this metric are given by
\bqn
H_1(t) &=& \sqrt{\frac{2 |\Lambda|}{3(\beta+2)}} \nb\\
&& \times \cot{\left( \sqrt{\frac{2 |\Lambda|}{3(\beta+2)}}  (t-t_0) + \sin^{-1}{\sqrt{\frac{|\Lambda|}{3}}} \right)} \,\,\,\,\,\\ 
\nb \\
H_2(t) &=& \sqrt{\frac{2 |\Lambda|}{3(\beta+2)}} \nb\\
&& \times \cot{\left( \sqrt{\frac{2 |\Lambda|}{3(\beta+2)}}  (t-t_0) - \sin^{-1}{\sqrt{\frac{|\Lambda|}{3}}} \right)} \,\,\,\,\,\\ 
\nb \\
q_1(t) &=&  \tan{\left( \sqrt{\frac{2 |\Lambda|}{3(\beta+2)}}  (t-t_0) + \sin^{-1}{\sqrt{\frac{|\Lambda|}{3}}} \right)^2} \\ 
\nb \\
q_2(t) &=& \tan{\left( \sqrt{\frac{2 |\Lambda|}{3(\beta+2)}}  (t-t_0) - \sin^{-1}{\sqrt{\frac{|\Lambda|}{3}}} \right)^2}
\eqn

The Kretschmann scalar for the metric is singular at
\bqn
t_{sing}(B_1) &=& t_0 - \sqrt{\frac{3(\beta+2)}{2 |\Lambda|}}   \sin^{-1}{\sqrt{\frac{|\Lambda|}{3}}}\\ 
\nb \\
t_{sing}(B_2) &=& t_0 + \sqrt{\frac{3(\beta+2)}{2 |\Lambda|}}   \sin^{-1}{\sqrt{\frac{|\Lambda|}{3}}}
\eqn
{For $\beta = 0$ the solution exists and is never singular with the curvature invariant being $\frac{8 |\Lambda|^2}{3}$. But, the metric is singular for $\beta+2 > 0$ {with} $\beta \neq 0$. This means it is a new singularity.}

The Raychaudhuri equation is given by
\bqn
\frac{d\theta}{d\tau}(B_1)&=& -\frac{ \lvert \Lambda \rvert }{2(\beta+2)}\times\nb \\ &&\frac{3\cos^2\left[\sqrt{\frac{2\lvert \Lambda \rvert}{3(\beta+2)}} (-t+C1)\right]-4}
{\cos^2\left[\sqrt{\frac{2\lvert \Lambda \rvert}{3(\beta+2)}}(-t+C1)\right]-1},
\lb{Lambda<0k-1B1}
\eqn
\bqn
\frac{d\theta}{d\tau}(B_2)&=& -\frac{ \lvert \Lambda \rvert }{2(\beta+2)}\times\nb \\ &&\frac{3\cos^2\left[\sqrt{\frac{2\lvert \Lambda \rvert}{3(\beta+2)}} (-t+C2)\right]-4}
{\cos^2\left[\sqrt{\frac{2\lvert \Lambda \rvert}{3(\beta+2)}}(-t+C2)\right]-1}.
\lb{Lambda<0k-1B2}
\eqn

\section{Conclusions}

The initial singularity in cosmology has been the center of much research even before the Big Bang model was vindicated by the COBE results in 1990 \cite{Mather:1991pc}. In fact, even before the advent of the singularity theorems, the issue of breakdown of predictability in classical physics at the spacetime singularities had particularly worried John Wheeler \cite{Wheeler1964}. There have been several efforts both classically and quantum mechanically to eliminate or avoid such singularities. See \cite{Struyve:2017jpl}, for example. 

In this paper, we investigated how the presence of timelike vector field in EA theory affects the nature of singularities in comparison to GR theory. For this, we first classified all the possible cosmological vacuum solutions in EA theory and found three cases which are non-singular in GR but are singular in EA theory. They are \textbf{A.3} ($\Lambda > 0$, $k = -1$), \textbf{B.3} ($\Lambda = 0$, $k = -1$) and \textbf{C.3} ($\Lambda < 0$, $k = -1$) which all are within the experimentally allowed parameter space satisfying $\beta+2>0$. Besides we found another singular solution in EA, \textbf{C.2} ($\Lambda<0, k=0$), which does not have a counterpart in GR, albeit it exists in EA only for $\beta+2 < 0$ which is ruled out by experiments.

Our conclusions are reinforced  by studying the focusing of congruence of timelike geodesics using the Raychaudhuri equation. 
For all the three singular cases \textbf{B.3} ($\Lambda=0, k=-1$), \textbf{C.2} ($\Lambda < 0, k=0$) and \textbf{C.3} ($\Lambda<0, k=-1$), the expansion rate is $\frac{d\theta}{d\tau} <  0$, assuring the convergence of the congruences. However, the convergence of the congruences of the case \textbf{A.3} ($\Lambda > 0$, $k = -1$) depends on the interval of time considered. For all the non-singular cases we have $\frac{d\theta}{d\tau} >  0$ as it should be. 

The important take away is that the new singular solutions show that GR and EA theories can have different global structures, even for the FLRW solutions when we go beyond flat geometry ($k=0$). The cases \textbf{B.1} and \textbf{B.3} with $\Lambda=0$  are particularly interesting since the vector field defining the preferred direction is the unique source of the curvature. This result challenges several preconceived ideas about the nature of spacetime singularities in modified theories of gravity.

\textcolor{black}{
Among the gravitational theories that break Lorentz invariance (LI) by construction, the most popular are Horava-Lifshitz (HL) theory and EA theory. In the case of HL theory, the breaking of LI is implemented by introducing a preferred foliation of space-time, but no additional structure. Whereas EA theory breaks LI by coupling general relativity to a dynamical unit timelike vector field. Jacobson \cite{Jacobson:2010mx} \cite{Jacobson:2013xta} showed that any hypersurface orthogonal solution to EA theory is a solution to the IR limit of a particular version of HL  gravity, although the converse does not appear to be true. For the purpose of the current paper, we would like to comment that the lowest dimension terms (the infrared limit (IR)) of the BPS \cite{Blas:2010} version of HL action are equivalent to those of EA theory when the aether vector is assumed to be hypersurface orthogonal. But, this does not mean that the EA theory is an IR theory of gravity. Unfortunately, there does not exist much literature on the ultraviolet (UV) behavior of EA theory. The second thing we would like to comment about is that we have confirmed the singularities purely from geometry without appealing to the energy conditions in EA theory, which are shown to be violated in some solutions \cite{Eling:2006df}. The third thing we would like to comment on is the role of breaking of LI on singularities. Perhaps the roots of EA theory go back to Gasperini \cite{Gasperini:1987nq} who suggested that in the UV,  the Lorentz gauge symmetry of general relativity could breakdown leading a way of evading the singularity theorems at least in a
classical geometric context. Although Gasperini showed the avoidance of singularities in FLRW universe with the matter by invoking repulsive gravitational interactions associated with a non-minimal breaking of the local Lorentz symmetry, that is not the case here.
}

\section*{Acknowledgments}

The financial assistance from FAPERJ/UERJ (MFAdaS) is gratefully acknowledged. The author (RC) acknowledges the financial support from FAPERJ (no.E-26/171.754/2000, E-26/171.533/2002 and E-26/170.951/2006). MFAdaS and RC also acknowledge the financial support from Conselho Nacional de Desenvolvimento Cient\'{\i}fico e Tecnol\'ogico - CNPq - Brazil.  The author (MFAdaS) also acknowledges the financial support from Financiadora de Estudos  Projetos - FINEP - Brazil (Ref. 2399/03). VHS gratefully acknowledges the financial support from FAPERJ though Programa P\'{o}s-doutorado Nota 10. VHS thanks Jos\'{e} Abdalla Helay\"{a}l-Neto for the hospitality at Centro Brasileiro de Pesquisas F\'{i}sicas (CBPF) and Anzhong Wang for many insightful discussions.


\begin{thebibliography}{200}


\bibitem{Crispino:2019yew} 
L.~C.~B.~Crispino and D.~Kennefick,
``100 years of the first experimental test of General Relativity,''
Nature Phys.\  {\bf 15}, 416 (2019)
[arXiv:1907.10687 [physics.hist-ph]].

\bibitem{Abbott:2016blz} 
B.~P.~Abbott {\it et al.} [LIGO Scientific and Virgo Collaborations],
``Observation of Gravitational Waves from a Binary Black Hole Merger,''
Phys.\ Rev.\ Lett.\  {\bf 116}, no. 6, 061102 (2016)
[arXiv:1602.03837 [gr-qc]].

\bibitem{Akiyama:2019cqa} 
K.~Akiyama {\it et al.} [Event Horizon Telescope Collaboration],
``First M87 Event Horizon Telescope Results. I. The Shadow of the Supermassive Black Hole,''
Astrophys.\ J.\  {\bf 875}, no. 1, L1 (2019)
[arXiv:1906.11238 [astro-ph.GA]].

\bibitem{Hawking:1969sw} 
S.~W.~Hawking and R.~Penrose,
``The Singularities of gravitational collapse and cosmology,''
Proc.\ Roy.\ Soc.\ Lond.\ A {\bf 314}, 529 (1970).

\bibitem{Will:2018bme} 
C.~M.~Will,
``Theory and Experiment in Gravitational Physics,''
(Cambridge University Press, 2018)

\bibitem{Campista:2018gfi} 
M.~Campista, R.~Chan, M.~F.~A.~da Silva, O.~Goldoni, V.~H.~Satheeshkumar and J.~F.~V.~da Rocha,
``Vacuum solutions in the Einstein-Aether Theory,''
[arXiv:1807.07553 [gr-qc].

\bibitem{Barrow:2004xh} 
J.~D.~Barrow,
``Sudden future singularities,''
Class.\ Quant.\ Grav.\  {\bf 21}, L79 (2004)
[gr-qc/0403084].

\bibitem{Barrow:2004hk} 
J.~D.~Barrow,
``More general sudden singularities,''
Class.\ Quant.\ Grav.\  {\bf 21}, 5619 (2004)
[gr-qc/0409062].

\bibitem{FernandezJambrina:2004yy} 
L.~Fernandez-Jambrina and R.~Lazkoz,
``Geodesic behaviour of sudden future singularities,''
Phys.\ Rev.\ D {\bf 70}, 121503 (2004)
[gr-qc/0410124].

\bibitem{Barrow:2015sga} 
J.~D.~Barrow and A.~A.~H.~Graham,
``New Singularities in Unexpected Places,''
Int.\ J.\ Mod.\ Phys.\ D {\bf 24}, no. 12, 1544012 (2015)
[arXiv:1505.04003 [gr-qc]].



\bibitem{Jacobson:2000xp} 
T.~Jacobson and D.~Mattingly,
``Gravity with a dynamical preferred frame,''
Phys.\ Rev.\ D {\bf 64}, 024028 (2001)
[arXiv:0007031 [gr-qc]].

\bibitem{Garfinkle}
D.~Garfinkle, C.~Eling and T.~Jacobson,
``Numerical simulations of gravitational collapse in Einstein-aether theory,''
Phys.\ Rev.\ D {\bf 76}, 024003 (2007)
[arXiv:0703093 [gr-qc]].

\bibitem{Jacobson:2008aj} 
T.~Jacobson,
``Einstein-aether gravity: A Status report,''
PoS QG {\bf -PH}, 020 (2007)
[arXiv:0801.1547 [gr-qc]].

\bibitem{Foster:2005dk} 
B.~Z.~Foster and T.~Jacobson,
``Post-Newtonian parameters and constraints on Einstein-aether theJacobson:2010mxory,''
Phys.\ Rev.\ D {\bf 73}, 064015 (2006)
[gr-qc/0509083].

\bibitem{Carroll:2004ai}
S.~M.~Carroll and E.~A.~Lim,
``Lorentz-violating vector fields slow the universe down,''
Phys.\ Rev.\ D {\bf 70}, 123525 (2004)
[arXiv:0407149 [hep-th]].

\bibitem{Elliott:2005va} 
J.~W.~Elliott, G.~D.~Moore and H.~Stoica,
JHEP {\bf 0508}, 066 (2005)
[arXiv:0505211 [hep-ph]]

\bibitem{Eling:2003rd} 
C.~Eling and T.~Jacobson,
Phys.\ Rev.\ D {\bf 69}, 064005 (2004)
[arXiv:0310044 [gr-qc]]

\bibitem{GrEAsser:2005bg} 
M.~L.~GrEAsser, A.~Jenkins and M.~B.~Wise,
Phys.\ Lett.\ B {\bf 613}, 5 (2005)
[arXiv:0501223 [hep-th]]

\bibitem{Foster:2007gr} 
B.~Z.~Foster,
Phys.\ Rev.\ D {\bf 76}, 084033 (2007)
[arXiv:0706.0704 [gr-qc]]

\bibitem{Yagi:2013ava} 
K.~Yagi, D.~Blas, E.~Barausse and N.~Yunes,
Phys.\ Rev.\ D {\bf 89}, no. 8, 084067 (2014)
Erratum: [Phys.\ Rev.\ D {\bf 90}, no.

\bibitem{Gong:2018cgj} 
Y.~Gong, S.~Hou, D.~Liang and E.~Papantonopoulos,
Phys.\ Rev.\ D {\bf 97}, 084040 (2018)
[arXiv:1801.03382]

\bibitem{Oost:2018tcv} 
J.~Oost, S.~Mukohyama and A.~Wang,
``Constraints on Einstein-aether theory after GW170817,''
Phys.\ Rev.\ D {\bf 97}, no. 12, 124023 (2018)
[arXiv:1802.04303 [gr-qc]].



\bibitem{Coley:2019tyx} 
A.~Coley and G.~Leon,
``Static Spherically Symmetric Einstein-aether models I: Perfect fluids with a linear equation of state and scalar fields with an exponential self-interacting potential,''
Gen.\ Rel.\ Grav.\  {\bf 51}, no. 9, 115 (2019)
[arXiv:1905.02003 [gr-qc]].

\bibitem{Leon:2019jnu} 
G.~Leon, A.~Coley and A.~Paliathanasis,
``Static Spherically Symmetric Einstein-aether models II: Integrability and the Modified Tolman-Oppenheimer-Volkoff approach,''
Annals Phys.\  {\bf 412}, 168002 (2020)
[arXiv:1906.05749 [gr-qc]].

\bibitem{Trinh:2018pcb} 
D.~Trinh, F.~Pace, R.~A.~Battye and B.~Bolliet,
``Cosmologically viable generalized Einstein-Aether theories,''
Phys.\ Rev.\ D {\bf 99}, no. 4, 043515 (2019)
[arXiv:1811.07805 [astro-ph.CO]].

\bibitem{Landau:1982dva}
L.~D.~Landau and E.~M.~Lifshitz,
``The Classical Theory of Fields,'' 
(Volume 2 of Course on Theoretical Physics Series) 4th Edition (Butterworth-Heinemann; 1980)

\bibitem{Albareti:2012se} 
F.~D.~Albareti, J.~A.~R.~Cembranos and A.~de la Cruz-Dombriz,
``Focusing of geodesic congruences in an accelerated expanding Universe,''
JCAP {\bf 1212}, 020 (2012)
[arXiv:1208.4201 [gr-qc]].

\bibitem{Raychaudhuri:1953yv} 
A.~Raychaudhuri,
``Relativistic cosmology. 1.,''
Phys.\ Rev.\  {\bf 98}, 1123 (1955).

\bibitem{Hawking:1973uf} 
S.~W.~Hawking and G.~F.~R.~Ellis,
``The Large Scale Structure of Space-Time,''
(Cambridge, 1973)

\bibitem{Wald:1984}
R.~M.~Wald,
``General Relativity,''
(University Chicago Press, 1984)
%
\bibitem{Senovilla:2006}  
J.~M.~M.~Senovilla,
``Singularity theorems in general relativity: Achievements and open questions,''
Einstein Stud.\  {\bf 12}, 305 (2012)
[physics/0605007].

\bibitem{d'Inverno1992} 
R. d'Inverno, ''Introducing Einstein's Relativity''(Oxford University Press, 1992) p. 340 




\bibitem{Mather:1991pc} 
J.~C.~Mather {\it et al.},
``A Preliminary measurement of the Cosmic Microwave Background spectrum by the Cosmic Background Explorer (COBE) satellite,''
Astrophys.\ J.\  {\bf 354}, L37 (1990).

\bibitem{Wheeler1964} 
J.A. Wheeler, 
``Geometrodynamics and the issue of the final state"
In \textit{Relativity, Groups and Topology - Lectures delivered at Les Houches during the 1963 session of the Summer School Of Theoretical Physics}, 
edited by C. deWitt and B. deWitt 
(Gordon and Breach, New York, 1964) pp. 317-522.

\bibitem{Struyve:2017jpl} 
W.~Struyve,
``Loop quantum cosmology and singularities,''
Sci.\ Rep.\  {\bf 7}, no. 1, 8161 (2017)
[arXiv:1703.10274 [gr-qc]].

\textcolor{black}{
\bibitem{Jacobson:2010mx}
  T.~Jacobson,
  ``Extended Horava gravity and Einstein-aether theory,''
  Phys.\ Rev.\ D {\bf 81}, 101502 (2010)
  Erratum: [Phys.\ Rev.\ D {\bf 82}, 129901 (2010)]
  [arXiv:1001.4823 [hep-th]].
%
\bibitem{Jacobson:2013xta}
  T.~Jacobson,
  ``Undoing the twist: The Hořava limit of Einstein-aether theory,''
  Phys.\ Rev.\ D {\bf 89}, 081501 (2014)
  [arXiv:1310.5115 [gr-qc]].
%
\bibitem{Blas:2010}
D. Blas, O. Pujolas, and S. Sibiryakov, Phys. Rev. Lett. 104, 181302 (2010) 
[arXiv:0909.3525]; JHEP, 04, 018 (2011) [arXiv.1007.3503].
%
\bibitem{Eling:2006df}
  C.~Eling and T.~Jacobson,
  ``Spherical solutions in Einstein-aether theory: Static aether and stars,''
  Class.\ Quant.\ Grav.\  {\bf 23}, 5625 (2006)
  Erratum: [Class.\ Quant.\ Grav.\  {\bf 27}, 049801 (2010)]
  [gr-qc/0603058].
%
\bibitem{Gasperini:1987nq}
  M.~Gasperini,
 ``Singularity Prevention and Broken Lorentz Symmetry,''
  Class.\ Quant.\ Grav.\  {\bf 4}, 485 (1987).
}

\end{thebibliography}
\end{document}